\documentclass[10pt,journal,compsoc,twocolumn]{IEEEtran}
\usepackage{times,amsmath,epsfig}
\usepackage[english]{babel}
\usepackage{amsmath,amsfonts,amsthm} 
\usepackage[hang, small, labelfont=bf, up, textfont=it]{caption}
\usepackage{booktabs} 
\usepackage{lastpage}
\usepackage{graphicx} 
 
\usepackage{enumitem}
\setlist{noitemsep} 
\usepackage{pstricks}
\usepackage{amssymb}
\usepackage[utf8]{inputenc}
\usepackage{hyperref}
\usepackage{multirow}
\usepackage{amsmath}
\usepackage{mathtools}
 \usepackage{tikz}
 \usepackage{bclogo}
 \usepackage{color, colortbl}
 \usepackage{amsthm}

\newtheorem{theorem}{Theorem}

\numberwithin{theorem}{section} 

\usepackage[ruled,vlined,linesnumbered]{algorithm2e}

\usepackage{geometry}
\usepackage[T1]{fontenc} 
\usepackage[utf8]{inputenc}

\usepackage{fancyhdr} 
\newcommand{\uprove}{{\textbf{U-prove }}}
\newcommand{\idemix}{{\textbf{Idemix }}}
\newcommand{\itwopa}{{\textbf{I2PA }}}

\newcommand{\idemixs}{{\textbf{Idemix's }}}

\newcommand{\uprovev}{{\textbf{U-prove, }}}
\newcommand{\idemixv}{{\textbf{Idemix, }}}

\newcommand{\uprovep}{{\textbf{U-prove. }}}
\newcommand{\idemixp}{{\textbf{Idemix. }}}
\newcommand{\itwopap}{{\textbf{I2PA. }}}

\title{I2PA: an efficient ABC for IoT}
\author{
{SENE Ibou{\small $~^{\#*1}$}, CISS Abdoul Aziz{\small $~^{\#2}$} and NIANG Oumar{\small $~^{\#3}$} }
\vspace{1.6mm}\\
\fontsize{10}{10}\selectfont\itshape
$^{\#}$\,Ecole Polytechnique de Thiès(E.P.T.), \\Laboratoire de Traitement de l’Information et des Systèmes Intelligents(LTISI),\\
PO Box A10, Thiès, Sénégal\\
\vspace{1.2mm}
\fontsize{10}{10}\selectfont\itshape
$^{*}$\,Université de Thiès(U.T.),\\ Ecole Doctorale Développement Durable et Société (ED2DS),\\
PO Box 967, Thiès, Sénégal\\
\fontsize{9}{9}\selectfont\ttfamily\upshape
$^{1}$\,senei@ept.sn,
$^{2}$\,aaciss@ept.sn,
$^{3}$\,oniang@ept.sn
}
\begin{document}
\maketitle

\begin{abstract} 
 Internet of Things (IoT) is very attractive because of its promises. However, it brings many challenges, mainly issues about privacy preserving and lightweight cryptography. Many schemes have been designed so far but none of them simultaneously takes into account these aspects. In this paper, we propose an efficient ABC scheme for IoT devices. We use ECC without pairing, blind signing and zero knowledge proof. Our scheme supports block signing, selective disclosure and randomization. It provides data minimization and transactions' unlinkability. Our construction is efficient since smaller key size can be used and computing time can be reduced. As a result, it is a suitable solution for IoT devices characterized by three major constraints namely low energy power, small storage capacity and low computing power.
\end{abstract}


\section{Introduction}
\label{label-introduction}
\hspace{3mm}
Internet has changed our way of living. In fact, it has become an integral part of our life. As a ubiquitous communication platform, it has undergone remarkable development in recent years. The concept of IoT has been emerged and envisages to integrate all real-world objects into the Internet. The evolution of IoT concept has given rise to others concepts such as \textbf{IoE} for Internet of Everything or \textbf{IoV} for Internet of Vulnerabilities. Some people even talk about \textbf{IoP}\cite{miranda2015internet} for Internet of People. According to Cisco forecasts\cite{ciscoprevision}, there would be 50 billions connected devices by 2020.

\vspace{1.5mm}In most cases, communications between devices require authentication itself based on authentication factors\cite{mbaye2014lightweight}. Authentication is also based on identification which makes activities of an entity traceable since each device is directly or indirectly associated with its owner. The current infrastructure is based on centralized architecture which allows no control of data by their owners; this is a threat to privacy preserving. Security in current infrastructures is guaranteed in most cases by PKIs (Public Key Infrastructures) that can guarantee  that messages are not compromised and only recipients are able to open and read them (integrity, confidentiality and authenticity are guaranteed). PKIs main objective is to guarantee keys encryption authenticity. However, they cannot protect users' privacy and users cannot get a credential on a subset of their attributes without letting the CA (Certification authority) see the resulting credential. Moreover, when authenticating to a service provider, users must show their whole credential instead of just proving their eligibility for that access. We must therefore think to migrate to decentralized architectures and user-centric.

\vspace{1.5mm}The ultimate challenges can be summarized to this fundamental question: \textbf{How to minimize the amount of data disclosed about oneself and provide only the bare minimum necessary ?} Nowadays, to the best of our knowledge, the most convenient way to protect users' privacy remains using \textbf{anonymous credential} systems also known as \textbf{Attribute-Based Credentials} (ABCs). ABCs are building blocks for user-centric identity management\cite{galparinproceedings}. With ABCs, it is  possible to get a signature on a set of attributes and then use this later to access others services. Seeing the abstractness of some attributes (nationality, age class, occupation, affiliation, profession, etc.), the entire mechanism can remain anonymous and therefore guarantees  concerned entities' privacy. Authentication is carried out by revealing the bare minimum necessary. Better, it is possible, for a set of attributes, to prove their possession instead of revealing their values; from where the concept \itwopa meaning \textbf{"I Prove Possession of Attributes".} 

\vspace{1.5mm}
There are two major families of ABC systems, namely those based on blind signatures (BS) and those based on zero-knowledge proofs (ZKP). In terms of anonymous credential, there are two flagship schemes; \idemix of IBM and \uprove of Microsoft\cite{paquin2011u}. While \idemixs building block\cite{camenisch2002design}  is based on Camenish-Lysyanskaya signature scheme \cite{camenisch2001efficient} (CL-Signature), \uprove is based on Stefan Brand's digital signature\cite{brands2000rethinking} instead. Nowadays, there are many contributions (\uprovev \idemixv IRMA of Radboud University of Nijmegen\cite{irma2016} based on \idemixv etc.). However, when tackling privacy concerns, none of this schemes is fully adapted in an IoT environment characterized by three major constraints namely low computing power, very limited storage capacity and low energy autonomy. In fact, these models are based either on RSA cryptosystem\cite{vullers2013efficient} or on Pairing-Based Cryptography\cite{camenisch2004signature}. Some of them are based on ECC without pairing (\uprove for instance) but do not fully take into account some fundamental features relative to privacy preserving. Indeed, Lucjan et al.\cite{hanzlik2014short} showed that unlinkability feature is not fully taken into account by \uprovep With reference to RSA based cryptosytems, the major problem remains keys' size which will necessarily rise problem of storage, performance in computing time and bandwidth usage. Pairing-based cryptosystems, although considered as very robust, are not applicable in an IoT context seeing that they are too greedy in term of computing time. The relative computation cost of a pairing is approximately twenty times higher than that of the scalar multiplication. The model presented by Gergely et al.\cite{alpar2012designated} is very efficient in a sensor network but does not guarantee anonymity since the verifier has a database of identifiers. Fuentes et al.\cite{de2017assessment} present a very interesting result in Vehicular ad hoc network but their scheme lacks generality. Even if \idemix is the most advanced in terms of implementation, it is subject of improvement. Authors in\cite{sinha2013performance} show that at equal security level, an RSA key is at least six times longer than an Elliptic Curve Cryptography (ECC) key. Further, in major cases, operations are faster on ECC than in RSA based schemes. Another important point is that, in an RSA cryptosystem, when keys' size are no longer sufficient to guarantee a secure level, it is recommended to double their size what is not necessarily the case for ECC cryptosystem. Another important contribution \cite{bernal2017holistic} is build on top of \idemix scheme. It has been successfully implemented, deployed, and tested. However, the criticisms of \idemix remain valid toward it. The \uprove scheme, by its generality, can be implemented either on the basis of subgroup or using ECC. However, as shown by Lucjan et al.\cite{hanzlik2014short}, unlinkability feature is not fully taken into account. Indeed, if a token is presented twice to a verifier, then the later knows that it is the same token. This means that unlinkability in \uprove can be achieved only by using different credentials\cite{alpar2015attribute}, what requires the client device to have additional storage space.

\vspace{1.5mm}The relevance of our contribution lies in the fact that it offers a good level of security and drastically reduces keys' size. Credentials' randomizations, selective disclosure and unlinkability features are very interesting results that contribute in guaranteeing non traceability. Edwards curves, known as the curves in which cryptographic calculations are faster\cite{liu2015performance}, are privileged. We then win in memory usage, performance in computing time and bandwidth usage. Analysis presented in section \ref{label-complexity} show that our scheme is very efficient. Its level of abstractness makes it applicable in any IoT environment.

\vspace{1.5mm}The rest of this paper is organized as flow. We start by reviewing some mathematic backgrounds in section \ref{label-background} while definitions of some flagship concepts are presented in section \ref{label-definition}. Section \ref{label-architecture} presents an architecture of ABC system and in section \ref{label-constribution}, our contribution is presented. Section \ref{label-complexity} is related to complexity analysis and this paper is ended by a conclusion and perspectives in section \ref{label-conclusion}.


\section{Mathematic background}
\label{label-background}
 \hspace{3mm}
Elliptic Curves Cryptography (ECC) was presented independently by Koblitz\cite{koblitz1987elliptic} and Miller\cite{miller1985use} in the 80s. Their structure of group and performance in computing time they offer make them a new direction in cryptography. The following section is a brief description on ECC.

\subsection{Définition}
An Elliptic Curve $\textbf{E}$ over a field $\mathbb{K}$ can be described as the set of $\mathbb{K}\times\mathbb{K}$ satisfying the equation :
\begin{equation}
E: y^2+a_1xy+a_3y=x^3+a_2x^2+a_4x+a_6
\end{equation}

where $a_i \in \mathbb{K}$ to which we add a point at infinity $ \cal O$, defined as the intersection of all vertical lines.
An additional requirement is that the curve must be "smooth", which means that the partial derivations $ \frac{\partial E}{ \partial y}$ and $ \frac{\partial E}{ \partial x}$ have no common zeros.
Depending on the characteristic of the field $\mathbb{K}$, the equation below can be simplified\cite{mbaye2014lightweight}\cite{aziz2015trends}.

\begin{enumerate}
\item {
When $char(\mathbb{K}) \ne 2,3$ the equation can be simplified to $y^2=x^3+ax+b$,~ where a, b $\in \mathbb{K} $. 
}
\item {When $char(\mathbb{K}) =2$ and $a_1\ne 0$, the equation can be simplified to $y^2+xy=x^3+ax^2+b$, where a, b $\in \mathbb{K}$. This curve is said to be non-supersingular. If $a_1=0$, the equation can be simplified to $y^2+cy=x^3+ax+b$,where a, b, c $\in \mathbb{K}$. This curve is said to be supersingular.
}
\item {
When $char(\mathbb{K}) =3$ and $a_{1}^{2}\ne -a_2$ the equation can be simplified to $y^2=x^3+ax^2+b$, where a, b $\in \mathbb{K} $. This curve is said to be non-supersingular. If $a_{1}^{2}= -a_2$, the equation can be simplified to
$y^2=x^3+ax+b,~a, b \in \mathbb{K} $. This curve is said to be supersingular.
}
\end{enumerate}

\begin{figure}[h] 
\center
\includegraphics[keepaspectratio=true,scale=0.4]{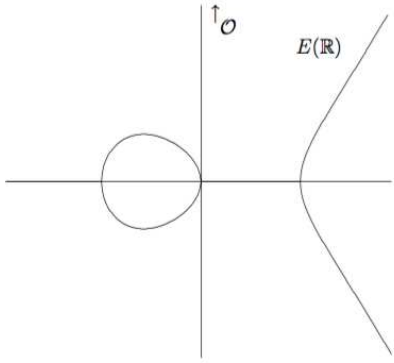} 
\caption{An Elliptic Curve of equation $y^2=x^3-x$ orver $\mathbb{R}$}
\label{fig:ecc-example}
\end{figure}

The "\textbf{figure} \ref{fig:ecc-example}" is an illustration of the curve $y^2=x^3-x$ over $\mathbb{R}$.

\begin{theorem}[Hasse’s theorem]
\label{hasstheorem}
Let $E$ be an elliptic curve defined over a finite field $\mathbb{K}$ with q elements, then the result of Hasse states that : 
$|\#E(\mathbb{K})-q-1| \le 2\sqrt{q}$
\end{theorem}

\subsection{Edwadrs' curves}
\hspace{3mm}
Edwards, generalizing an example from Euler and Gauss, introduced an addition law for the curves $x^2+y^2=c^2(1+x^2y^2)$ over a non-binary field $\mathbb{K}$. He showed that every elliptic curve over a non-binary field $\mathbb{K}$ can be expressed in the form $x^2+y^2=c^2(1+x^2y^2)$ if $\mathbb{K}$ is algebraically closed. Bernstein and Lange\cite{bernstein2007faster} generalized the addition law of the curves $x^2+y^2=c^2(1+dx^2y^2)$. Let $char(\mathbb{K}) \ne 2,3$ and let $E(\mathbb{K})$ has a unique point of order 2. Then, $E$ can be written in Edwards form :

\begin{equation}
E_d: x^2+y^2=1+dx^2y^2, where ~d \notin \{ 0,1\}
\end{equation}

Let $P_1(x_1,y_1) $, $P_2(x_2,y_2)$ and $P_3(x_3,y_3)$ be points of the curve such that $P_3=P_1\bigoplus P_2$. The Edwards addition law over \textbf{$E_d$} is described as follow :
\begin{equation*}
(x_3,y_3)=(\frac{x_1y_2+x_2y_1}{1+dx_1x_2y_1y_2},\frac{y_1y_2-x_1x_2}{1-dx_1x_2y_1y_2})
\end{equation*}
This group law is complete and strongly uniform. The doubling operation is given by the following equation :
\begin{equation*}
2(x,y)=(\frac{2xy}{x^2+y^2} , \frac{y^2-x^2}{2-x^2-y^2})
\end{equation*}
This addition law presents interesting results :

\begin{itemize}
\item{If d is a non-square in $\mathbb{K}$, the addition law is complete. This ensures that denominators are never zero.}
\item{The addition law is strongly unified, i.e., it can be also used for doubling.}
\item {The point (0,1) is the neutral element. }
\item {The point (0,-1) has order 2. }
\item {The points ($\pm$1,0) have order 4. }
\item {The inverse of $(x, y)$ is $(-x, y)$. }
\end{itemize}

\vspace{1.5mm}
Addition in Edwards curve is known to be the fastest among all families of elliptic curves used to implement Cryptography.

\subsection{Elliptic Curve Discret Logarithm Problem (ECDLP)}
The basic operations of ECC are point scalar computations (also known as scalar multiplication) of the form :
\[
Q=k.P=\underbrace{P+P+\ldots +P}_{\text{k times}}
\]

The Elliptic Curve Discrete Logarithm Problem (ECDLP) is the problem of retrieving k given P and Q, where P and Q are points of the curve and k is an integer uniformly chosen at random in the interval [1,r-1](r denotes the order of P). The assumed difficulty of this problem is the basis of security in elliptic curve public key cryptosystems. Point scalar multiplication can be performed efficiently using algorithms such as  Double-and-Add .

\section{Definition}
\label{label-definition}
\hspace{3mm}
The aim of this section is to define some leading terms in this paper namely attribute, credential, blind signing and zero-knowledge-proof.
\label{label-definition}

\subsection{Attribute}
\label{label-definition-attribute}
\hspace{3mm}
An attribute is a characteristic or a qualification of a person. It can either be an identifying or non-identifying property. For example, \textbf{"full name"}, \textbf{"address"}, \textbf{"social security number"}, are identifying attributes. Attributes such as \textbf{"is a student"} and \textbf{"is a teenager"} are non-identifying as they do not uniquely identify a person; such properties can belong to other people as well. 

\subsection{Credential}
\label{label-definition-credential}
\begin{figure}[h] 
\center
\includegraphics[keepaspectratio=true,scale=0.49]{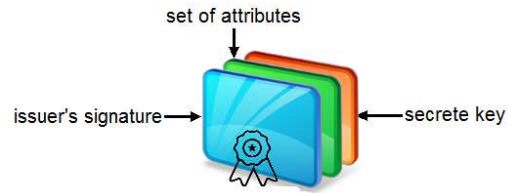} 
\caption{Credential's structure}
\label{fig:credential}
\end{figure}

\hspace{3mm}
A credential is a set of attributes digitally signed by a trusted third party(issuer). In others words, a set of attributes together with the corresponding cryptographic information. A credential is similar to a certificate in terms of content but they are different in the way they are used. While certificate usage requires showing all its content, a credential can be used by showing some parts and hiding or proving knowledge of others. Credentials are important in identity management systems. They certify that an entity has certain characteristics, knowledges, skills, etc. One main point is that credentials involve attributes of an entity without including identity information which allows linking the credential to its owner. As illustrated in the "\textbf{figure} \ref{fig:credential}", a credential has three main parts: a secret key of its owner, a set of attributes and a signature of an issuer on that set of attributes. A credential includes others informations such as expiry date.

\subsection{Zero-Knowledge Proofs}
\label{label-definition-zkp}
\hspace{3mm}
The past decades have witnessed the emergence of several new cryptographic notions. In 1985, Goldwasser, Micali and Rackoff \cite{goldwasserMR85} introduced the concept of zero-knowledge interactive proofs that enables an entity, a \textbf{prover}, to convince another entity, a \textbf{verifier}, of the validity of a statement without revealing anything else beyond the assertion of this statement. Zero-Knowledge Proofs are elegant techniques to limit the amount of information transferred from a prover to a verifier in a cryptographic protocol. 
%
The "\textbf{table} \ref{label-table-zkp}" describes Schnorr's proof of knowledge also known as Schnorr's Identification protocol. Given a group G of prime order q, in which the discrete logarithm problem (DLP) is hard, and a generator g ($\langle g \rangle = G$), a prover proves to a verifier that he knows a secret value $x$, uniformly chosen at random from $ \mathbb{Z}_{q}^{*}$, corresponding to a public value $h = g^x \in G$.
\begin{center} 
\captionof{table}{Schnorr's proof of knowledge }
\label{label-table-zkp}
\begin{tabular}{|l|c|r|}
\hline 
\multicolumn{1}{|p{3cm}|}{\centering \textbf{Prover } \\ Secret $x$} 
& \multicolumn{1}{|p{1.5cm}|}{\centering \textbf{Public } \\ $ q,g,h=g^x$} 
&\multicolumn{1}{|p{2cm}|}{\centering \textbf{Verifier }} \\
\hline
\hline
\multicolumn{1}{|p{3cm}|}
{\centering 

$ w \in _{R} \mathbb{Z}_{q}^{*}$ \\ 
$a=g^{w}$ in G \\ 
~\\ 
$c =cx+w (mod ~q)$ \\
} 
& 
\multicolumn{1}{|p{1.5cm}|}
{\centering 
~\\ 
$ \overset{a}{ \xrightarrow{\hspace*{1cm}}} $ \\ 
$ \overset{c}{ \xleftarrow{\hspace*{1cm}}} $ \\ 
$ \overset{r}{ \xrightarrow{\hspace*{1cm}}} $ \\ 
} 
&
\multicolumn{1}{|p{2cm}|}
{\centering 
~\\ 
~\\ 
$c\in _{R} \mathbb{Z}_{q}$
~\\ 
$a \overset{?}{=} g^{r}.h^{-c}$ in G \\
} 
\\
\hline
\end{tabular}
\end{center}

\vspace{1.5mm}
There is a formal symbolic notation of ZKP  as described in "\textbf{equation} \ref{label:zkp-notation}".
\begin{equation}
\label{label:zkp-notation}
PK\{(\alpha): h=g^{\alpha}\}
\end{equation}

\subsection{Blind signing}
\label{label-definition-blind-signing}
\hspace{3mm}
In many applications involving anonymity, it is often desirable to allow a participant to sign a document without knowing its content; this is known as blind signature. It is typically used in privacy-related protocols where the signer and credential owner are different parties. It can be used in e-cash or privacy-vote. 

\subsection{Blindness}
\label{label-blindness}
\hspace{3mm}
Let $U_0$ and $U_1$ be two honest users and $A$ be a PPT (Probabilistic Polynomial-Time) adversary which plays the role of the signer engaged in the issuing scheme with $U_0$ and $U_1$ on messages $m_b$ and $m_{1-b}$ where $b$ is chosen uniformly at random in $\{0,1\}$. $U_0$ and $U_1$ output the signatures $\sigma_b$ and $\sigma_{1-b}$. After that,  $(m_b, m_{1-b}, \sigma_b, \sigma_{1-b})$ is sent to $A$ which outputs $b'\in\{0,1\}$. For all $A$, $U_0$, $U_1$, constant $c$, large $n$, we have $|\mathrm{Pr}[b=b']|< n^{-c}$.

\section{Architecture}
\label{label-architecture}
\begin{figure}[h] 
\center
\includegraphics[keepaspectratio=true,scale=0.5]{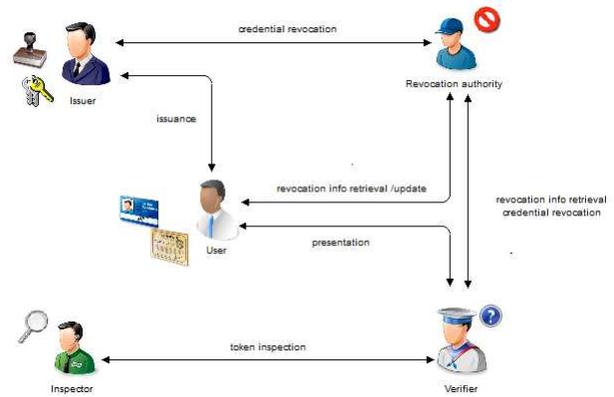} 
\caption{General architecture of ABC system}
\label{fig:abs-architecture}
\end{figure}

\hspace{3mm}
A general architecture of Privacy-ABC  consists of three main entities,  a User (credential owner also known as Prover), an Issuer (trusted third party or credential signer ) and a Verifier (services provider or secure resource owner). The Issuer issues credential(s) for a User, which can later be used for authentication purposes. Such an architecture may optionally include an entity that takes care of revocation of credentials (Revocation authority) and another entity (Inspector) that can revoke the anonymity of users. Inspection and revocation are beyond the scope of this paper. The main phases are described in the following section.
\begin{itemize}
\item {\textbf{Set-up :} Performed to output system's parameters}
\item {\textbf{Issuance :} An interactive protocol between a User and an Issuer. By issuing a credential to a User, the Issuer guarantees the correctness of attributes' values contained in the credential.}
\item {\textbf{Presentation :} An interactive protocol in which a User reveals or proves possession of some attributes or claims about attributes. This phase is also known as verification.}
\item {\textbf{Inspection :} Provides conditional anonymity. It enables a trusted party, the so called Inspector, to revoke, in some conditions, anonymity of cheating provers.}
\item {\textbf{Revocation :} Ends the validity of credentials whenever necessary.
}
\vspace{1.5mm}
\end{itemize}
All those phases are illustrated in "\textbf{figure} \ref{fig:abs-architecture}".

At the core of Privacy-ABCs systems, \textbf{untraceability} and \textbf{unlinkability} are the  most important privacy-related features. Additional features are also supported\cite{alpar2015attribute}\cite{galparinproceedings}, without exhaustivity, we cite.

\begin{itemize}

\item {\textbf{Authenticity :} Refers to the feature that guarantees that the content of an ABC signed by the issuer cannot be modified 
}

\item {\textbf{Non-transferability :} Refers to the  feature that guarantees that  prevents the user from transferring her ABC to another user of the system. 
}

\item {\textbf{Minimal information :} Refers to the privacy feature that guarantees that during the verification protocols no other information is revealed to the verifier beyond the disclosed attributes, the credential names and the corresponding issuers.}

\item {\textbf{Multi-show unlinkability :} Refers to the privacy feature that guarantees that different presentations of a given credential cannot be linked. 
}
\item {\textbf{Issuance unlinkability :} Refers to the privacy feature which guarantees that the presentation of a credential cannot be linked to its issuance.
}
\item {\textbf{Selective disclosure :} Allows a user to prove only a subset of attributes to a verifier. 
}
\item {\textbf{Carry-over attributes :} Enables users to carry over some attributes from an existing credential into a new one without disclosing them to
the Issuer.
}
\item {\textbf{Predicate proof :}  Allows logical operators, such as greater or smaller than, to be applied on attributes without disclosing them;
}
\item {\textbf{Prove of holdership :} A cryptographic evidence for proving ownership or possession of a credential.
}

\item {\textbf{Unforgeability :} Refers to the feature that guarantees that none malicious third party can forge a valid ABC.
}
\item {\textbf{Etc.} }
\end{itemize}

\section{Contribution}
\label{label-constribution}
\hspace{3mm}
In this sections, our scheme is described. We focus on three main phases namely set-up, issuance and verification (showing, presentation). We also present a selective disclosure protocol and randomization of credential.


\subsection{Set-up}
\label{label-set-up}
\hspace{3mm}
The set-up is the first algorithm to be ran to extract system's parameters from a security parameter k. This algorithm is described as follows :

\begin{itemize}
\item{ {\Large $\{p,E(\mathbb{F}_{p}), P, P_{pub}, \cal {H}$ $\} \longleftarrow 1^{k}$} }
\end{itemize}
Where :
\begin{itemize}

\item{ k :  a security parameter }
\item{p :  a prime number that defines the field $\mathbb{F}_{p}$.
}
\item{E : an elliptic curve defined over $\mathbb{F}_{p}$.}
\item{$P \in E(\mathbb{F}_{p})$ : a base point of prime order q.
}
\item{$x \in_{R} \mathbb{F}_{q}^{*}$ : issuer's secret key.
}
\item{$P_{pub}=x.P$ :  issuer's public key. }

\item{$\cal {H}$ :  a hash function defined as follow : 
\begin{align*}
 \cal{H} : & E(\mathbb{F}_{p})^{2} \xrightarrow{\hspace*{0.7cm}} \mathbb{F}_{p}^{*}\\
 & (P,Q) \xmapsto{\phantom{L^\infty(T)}} {\cal {H}}{(P,Q) }
\end{align*}
}
\end{itemize}

\subsection{Issuance}
\label{label-issuance}
\hspace{3mm}
When the set-up algorithm is successfully ran,  the system's parameters are available and ready to be used. The issuer is ready to issue credentials. When a user wants to be issued a credential, he runs an interactive algorithm with the issuer at the end of which a credential should be issued for him. Issuance unlinkability property should be ensured. The way it is done is that not only the issuer can not store the issued credential but also, thanks to the blinding  mechanism, he may not see the credential's content while signing. 
The issuing protocol on a single message is an interactive protocol of three steps between Prover and Issuer as described below : 

 \begin{itemize}
\item {\textbf{Blinding :} 
The issuer starts by generating a random integer $\bar{k} \in \mathbb{F}_{q}^{*}$, computes the resulting point $\bar{R}=\bar{k}.P$ and sends $\bar{R}$ to the User.  After receiving $\bar{R}$, the User generates random factors $\alpha$ and $\beta $. He blinds his document and sends it to the Issuer. The User should prove knowledge of $m_0$, this is very important to guarantee unforgeability.
}
\item {\textbf{Singing :}
After receiving the blinded document, the Issuer signs it with his secret key $x$ by computing $\bar{s} \equiv \bar{h}x+\bar{k} (mod ~q) $. He sends the blinded and signed document to the User.
}
\item {\textbf{Unblinding :}
After receiving the blinded and signed document, the User unblinds it with his blind factors $\alpha$ and $\beta$ without invalidating it. He can also verify that the signature is correctly computed.
}
\end{itemize}

\subsection{Signature on a single message}
\label{label-issuance-single-message}
\hspace{3mm}
For convenient notation, all proofs of knowledge would be noted $PK$. The corresponding proofs is those mentioned in the so called protocol. All the details of the issuing protocol on a single message are described in "\textbf{table} \ref{label-table-issuance-single-message}".

\begin{center} 
\captionof{table}{Issuance protocol on a single message}
\label{label-table-issuance-single-message}
\begin{tabular}{|l|c|r|}
\hline 
\multicolumn{1}{|p{3cm}|}{\centering \textbf{User }:$s_k$} 
& \multicolumn{1}{|p{1.6cm}|}{\centering \textbf{Public } :$ p_k$} 
& \multicolumn{1}{|p{2cm}|}{\centering \textbf{Issuer } : $x$ } \\
\hline
 \hline
\multicolumn{1}{|p{3cm}|}
{\centering 
\underline{Blinding}\\
$ \alpha, \beta \in _{R} \mathbb{F}_{q}^{*}$ \\ 
$R = \alpha. \bar{R}+\beta.P $ \\ 
$P_{0}=m_{0}.P $ \\ 
$h = {\cal {H}}(P_{0},R)$ \\ 

$\bar{h} \equiv h\alpha^{-1} (mod ~q)$ \\
$PK\{(\mu):P_{0}=\mu.P\}$\\
\underline{Verification}\\
$\bar{s}.P \overset{?}{=} \bar{h}.P_{pub}+\bar{R}$\\ 
\underline{Unblinding}\\
$s \equiv \alpha \bar{s}+\beta (mod ~q)$ \\
Outputs (R,s) the blind version of $ (\bar{R},\bar{s})$
} 
& 
\multicolumn{1}{|p{1.6cm}|}
{\centering 
~\\ 
$ \overset{\bar{R}}{ \xleftarrow{\hspace*{1.3cm}}} $\\ 
~\\ 
~\\ 
~\\ 
~\\ 
$ \overset{\bar{h}, P_{0}, PK}{ \xrightarrow{\hspace*{1.3cm}}} $\\ 

$ \overset{\bar{s}}{ \xleftarrow{\hspace*{1.3cm}}} $\\ 
} 
&
\multicolumn{1}{|p{2cm}|}
{\centering 
~\\ 
$\bar{k} \in _{R} \mathbb{F}_{q}^{*},Ò\bar{R}=\bar{k}.P$
~\\ 
~\\ 
~\\ 
~\\ 
\underline{Signing}\\
$\bar{s} \equiv \bar{h}x+\bar{k} (mod ~q) $
} 
\\
\hline
\end{tabular}
\end{center} 

\begin{theorem}
\label{label-theorem-blind-signing}
The prosed scheme is fully blind.
\begin{proof}[Proof]
From $\bar{s} \equiv \bar{h}x + \bar{k} (\mathrm{mod}\ q)$, we have $\alpha \equiv hx(\bar{s}-\bar{k})^{-1} (\mathrm{mod}\ q)$. Thus $\alpha \in \mathbb{F}_q^{*}$ is unique. It follows that $\beta \equiv s - \alpha\bar{s} (\mathrm{mod}\ q)$ is also unique. Thus, there always exist $\alpha$ and $\beta$, regardless of $(\bar{R}, \bar{h}, \bar{s})$ and $(m, R, s)$, such that $(\bar{R}, \bar{h}, \bar{s})$ and $(m, R, s)$ have the same relation. Therefore, an adversary $A$ outputs a correct value $b'$ with probability exactly $\frac{1}{2}$. As a result, the  issuing protocol is fully blind.
\end{proof}
\end{theorem}

\begin{theorem}
\label{label-theorem-resist-attacks}
The proposed scheme is $(\epsilon^{ \prime} ,t^{ \prime},q_{i},q_{h})$-secure in the sense of unforgeability under chosen message attack (UE-CMA) in the random oracle model, assuming that the $(\epsilon,t)$-ECDL assumption holds in $\langle P \rangle $, where $t^{\prime} =t +{\cal O}(q_{i})T $ , 
 $ \epsilon^{\prime}=(1-\frac{q_hq_i}{q})(1-\frac{1}{q})(\frac{1}{q_h})\epsilon$ and $q_{i},q_{h}$ are the number of issue and hashing queries, respectively, the adversary is allowed to perform while T denotes the time for a scalar multiplication operation.

\begin{proof}[Proof]
Assuming that there exists a forger $\mathcal{A}$ that can forge a credential while playing the game of chosen message attack (CMA)\cite{romahulsing2016}, we construct an algorithm $\mathcal{M^{A}}$ that uses $\mathcal{A}$ to solve the discrete logarithm problem. Without losing generality, we assume that $q_iq_h<q$. The following section is adapted from the proof presented by Joseph K Lui et al\cite{liu2010efficient}.
\begin{itemize}
\item {
\textbf{Setup}: $\mathcal{M^{A}}$ receives the problem $	P_1: (k, p, q, E(\mathbb{F}_{p}), P, P_{sec})$ and should find $x$ such that $P_{sec}=x.P$. It chooses a hash function $\mathcal{H}$ which behaves like a random oracle, sets $P_{pub}= P_{sec}$ and sends the public parameters $(k, p, q, E(\mathbb{F}_{p}), P, P_{pub}, \mathcal{H})$ to $\mathcal{A}$ expecting it forges a credential. $\mathcal{M^{A}}$ and $\mathcal{A}$ start playing the game of chosen message attack\cite{romahulsing2016}. 
}
\item {
\textbf{Hashing oracle}: $\mathcal{M^{A}}$ starts by initializing and empty database. When $\mathcal{A}$ sends $m_i$ for hashing, $\mathcal{M^{A}}$ checks  whether  or not that message has already been sent. If so, it picks $h_i$ from database and returns it as a response to that query, otherwise it picks $h_i$  uniformly at random from $\mathbb{F}_{p}^{*}$ ,stores the couple $(m_i,h_i)$ in the database and returns $h_i$ as a response to that query.
}
\item {
\textbf{Issuing oracle}: When $\mathcal{A}$ queries the issuing oracle for the message $m_i$, $\mathcal{M^{A}}$ first checks whether $m_i$ has already been queried for issuing. If so, it aborts and the game is stopped (\textbf{Event 1})\cite{romahulsing2016}, otherwise, it computes $h_i=RO(m_i)$(where $RO$ denotes the hashing oracle), picks random value $s_i$ from $\mathbb{F}_{p}^{*}$, computes $R_i=s_i.P-h_i.P_{pub}$ and sends $(R_i,s_i)$ to $\mathcal{A}$ as response to its issuing query. As we can see, the algorithm is valid since $s_i.P=h_i.P_{pub}+R_i$.
}
\item {
\textbf{Forging step}: Finally $\mathcal{A}$ outputs a forged signature $\sigma^{*}=(s^{*}, R^{*})$ on message $m^{*}$ with $h^{*}$. It computes $\frac{s^{*}-r^{*}}{h^{*}}$ and extracts $x$ as solution of the DLP.
}
\item {
\textbf{Probability analysis}: The simulation fails if $\mathcal{A}$ queries the same message for issuing (\textbf{Event 1}). This happens with probabilité at most ${(\frac{q_h}{q})}$. Hence, the simulation is successful with probability at least ${(1-\frac{q_h}{q})}^{q_i}\ge{(1-\frac{q_hq_i}{q})}$ (provable by recurrence reasoning). The tuple $({m^*},{R^*},{s^*})$ is a valid credential with probability at least $1-\frac{1}{q}$ and $\mathcal{M^{A}}$ guesses it correctly with probability at least $\frac{1}{q_h}$\cite{liu2010efficient}. Finally, the overall successful probability is $\epsilon^{\prime}=(1-\frac{q_hq_i}{q})(1-\frac{1}{q})(\frac{1}{q_h})\epsilon$\cite{liu2010efficient}. The time complexity of the algorithm $\mathcal{M^{A}}$ is $t^{\prime}=t + {\cal O}(q_{i})T$ since the issuing oracle computes at most $2q_iT$ scalar multiplications.
}
\end{itemize}

\end{proof}
\end{theorem}

\subsection{Verification}
\label{label-verification}
\hspace{3mm}
Once the credential is issued, the user can be authenticated by a service provider. As for the issuing algorithm, he must perform an interactive protocol with the service provider at the end of which an access to the so called service might be granted or denied. The prover must prove knowledge of his secret key. This guarantees unforgeability property. The verifying protocol is described in "\textbf{table} \ref{label-table-verification}".
\begin{center} 
\captionof{table}{Verification protocol}
\label{label-table-verification}
\begin{tabular}{|l|c|r|}
\hline 
\multicolumn{1}{|p{3cm}|}{\centering \textbf{Prover } :$s_k$ } 
& \multicolumn{1}{|p{1.6cm}|}{\centering \textbf{Public }:$p_k$} 
& \multicolumn{1}{|p{2cm}|}{\centering \textbf{Verifier } } \\
\hline
 \hline
\multicolumn{1}{|p{3cm}|}
{\centering 
Keeps secret $ (\bar{R},\bar{s})$ \\
$PK\{(\mu):P_{0}=\mu.P\}$
} 
& 
\multicolumn{1}{|p{1.6cm}|}
{\centering 
~\\ 
$ \overset{(R,s), h,PK}{ \xrightarrow{\hspace*{1.3cm}}} $ 
} 
&
\multicolumn{1}{|p{2cm}|}
{\centering 
$s.P \overset{?}{=} h.P_{pub}+R$ 
} 
\\
\hline
\end{tabular}
\end{center}

\subsection{Randomized version}
\label{label-randomization}
\hspace{3mm}
With rapid development of modern technologies, many digital services involved in our life emphasize user privacy. Blind signature is a well-known technique to address privacy concerns. 
When a user presents the same signature $(R, s)$ multiple times, he could be traceable; R, s could be stored by the verifier even if this can not be linked to issuance. The randomized version allows a user to derive a random signature from a valid one without invalidating it. The randomization feature is fundamental because it contributes in guaranteeing multi-show unlinkability. The prover generates a random factor $r$ from which a random signature $(\hat{R},\hat{s}) $ is derived. The randomization process is described in "\textbf{table} \ref{label-table-randomization}".

\begin{center} 
\captionof{table}{Randomization protocol}
\label{label-table-randomization}
\begin{tabular}{|l|c|r|}
\hline 
\multicolumn{1}{|p{3cm}|}{\centering \textbf{Prover } :$s_k$ } 
& \multicolumn{1}{|p{1.6cm}|}{\centering \textbf{Public }:$p_k$
} 
& \multicolumn{1}{|p{2cm}|}{\centering \textbf{Verifier } } \\
\hline
 \hline
\multicolumn{1}{|p{3cm}|}
{\centering 
$r \in _{R} \mathbb{F}_{q}^{*}$ \\ 
$\hat{s} \equiv s+r (mod ~q)$ \\ 
$\hat{R} =R+r.P$ 
} 
& 
\multicolumn{1}{|p{1.6cm}|}
{\centering 
~\\ 
$ \overset{(\hat{R},\hat{s}),h, PK}{ \xrightarrow{\hspace*{1.3cm}}} $ 
} 
&
\multicolumn{1}{|p{2cm}|}
{\centering 
~\\ 
$\hat{s}.P \overset{?}{=} h.P_{pub}+\hat{R}$ 
} 
\\
\hline
\end{tabular}
\end{center}  

\subsection{Signature on a block of messages}
\label{label-issuance-block-of-message}
\hspace{3mm}
A credential rarely contains one attribute. In  this section, we consider a credential of $l$ attributes $(m_1,...,m_l)$. The issuing protocol on a block of messages is described in  "\textbf{table} \ref{label-table-issuance-block-of-message}" . The verification protocol remains the same as in section \ref{label-verification}.

\begin{center} 
\captionof{table}{Issuance protocol on a block of messages}
\label{label-table-issuance-block-of-message}
\begin{tabular}{|l|c|r|}
\hline 
\multicolumn{1}{|p{3cm}|}{\centering \textbf{User }: $s_k$} 
& \multicolumn{1}{|p{1.55cm}|}{\centering \textbf{Public }:$p_k$} 
& \multicolumn{1}{|p{2cm}|}{\centering \textbf{Issuer}: $x$ } \\
\hline
 \hline
\multicolumn{1}{|p{3cm}|}
{\centering 
\underline{Blinding}\\
$ \alpha, \beta \in _{R} \mathbb{F}_{q}^{*}$ \\ 
$R \equiv \alpha. \bar{R}+\beta.P $ \\ 
$h = \prod_{i=0}^{l} {{\cal {H}}(P_{i},R)}$ \\ {\color{black} Where $P_{i}=m_{i}.P $} \\ 
$\bar{h} \equiv h\alpha^{-1} (mod ~q)$ \\
$PK\{(\mu):P_{0}=\mu.P\}$\\
\underline{Verification}\\
$\bar{s}.P \overset{?}{=} \bar{h}.P_{pub}+\bar{R}$\\ 
\underline{Unblinding}\\
$s \equiv \alpha \bar{s}+\beta (mod ~q)$ \\
Outputs (R,s) the blind version of $ (\bar{R},\bar{s})$
} 
& 
\multicolumn{1}{|p{1.55cm}|}
{\centering 
~\\ 
$ \overset{\bar{R}}{ \xleftarrow{\hspace*{1.3cm}}} $\\ 
~\\ 
~\\ 
~\\ 
~\\ 
$ \overset{\bar{h}, P_{0}, PK}{ \xrightarrow{\hspace*{1.3cm}}} $\\ 

$ \overset{\bar{s}}{ \xleftarrow{\hspace*{1.3cm}}} $\\ 
} 
&
\multicolumn{1}{|p{2cm}|}
{\centering 
~\\ 
$\bar{k} \in _{R} \mathbb{F}_{q}^{*},Ò\bar{R}=\bar{k}.P$
~\\ 
~\\ 
~\\ 
~\\ 
\underline{Signing}\\
$\bar{s} \equiv \bar{h}x+\bar{k} (mod ~q) $
} 
\\
\hline
\end{tabular}
\end{center} 

\subsection{Selective disclosure}
\label{label-selective-disclosure}
\hspace{3mm}
 The fundamental principle of privacy master is data minimization. Selective disclosure is a way to achieve this. It is the ability of an individual to granularly decide what information to share. In our context , it is a very interesting feature that lets a user  decides what attributes to disclose while being authenticated. How user and verify agreed on the attributes to disclose is beyond the scope of this paper. After convinced about hidden and revealed attributes, the verifier grants access to the prover. More concretely, in the selective disclosure protocol, the prover decides to disclose a subset of  attributes, let us say $m_{n+1},m_{n+2},...,m_{n+l}$ while $m_0,m_1,...,m_n$ remain secret.  The selective disclosure protocol is described in "\textbf{table} \ref{label-table-selective-disclosure}".

\begin{center} 
\captionof{table}{Selective disclosure protocol}
\label{label-table-selective-disclosure}
\begin{tabular}{|l|c|r|}
\hline 
\multicolumn{1}{|p{4cm}|}{\centering \textbf{Prover }: $m_0,m_1,...,m_n$ } 
& \multicolumn{1}{|p{1.3cm}|}{\centering \textbf{Public } :$ p_k$} 
& \multicolumn{1}{|p{1.3cm}|}{\centering \textbf{Verifier } } \\
\hline
 \hline
\multicolumn{1}{|p{4cm}|}
{\centering 
$PK\{ (s,R,\mu_{0},...,\mu_{n}):
\prod_{i=0}^{n} {{\cal {H}}(P_{i},R).P_{pub}}=(
\prod_{i=n+1}^{l} {{\cal {H}}(P_{i},R)})^{-1}(s.P-R) \wedge$ \\ $P_{i}= \mu_{i}.P,~0 \le i \le n$ \} 
} 
& 
\multicolumn{1}{|p{1.3cm}|}
{\centering 
~\\ 
$ \overset{(R,s),h, PK}{ \xrightarrow{\hspace*{1.2cm}}} $ \\ 
} 
&
\multicolumn{1}{|p{1.3cm}|}
{\centering 
~\\ 
$s.P \overset{?}{=} h.P_{pub}+R$ \\
} 
\\
\hline
\end{tabular}
\end{center} 


\section{Complexity analysis}
\label{label-complexity}
\hspace{3mm}
Design an algorithm is good, design an efficient algorithm is better. 
Let P be a problem, $M_1$ and $M_2$ two methods designed to solve P. Answer to the question: which of the methods $M_1$ and $M_2$ is more efficient is not trivial unless one knows their complexity. A complexity is a mathematical approximation to estimate the number of operations and/or memory required for an algorithm to solve a problem. A good algorithm must therefore use as little memory as possible and make the processor work less than possible. In the remain of this section, we adopt the following notations :

\begin{itemize}
\item {$M_s$ : Scalar multiplication over EC}
\item {$A_p$ : Point adding over EC }
\item {$I$ : Inversion }
\item {$A$ : Addition }
\item {$M$ : Multiplication }
\item {$P$ : Elevation to power }
\end{itemize}

\hspace{3mm}This section aims  to compare complexities of schemes  \idemixv  \uprove and \itwopap 
While tackling IoT devices, the case of \idemix could be left behind because a basic RSA operation (inversion or elevation to power on large numbers) is more expensive than a well designed operation on an EC\cite{sinha2013performance}\cite{jansma2004performance}. It should be noted that basic arithmetic operations such as addition and multiplication are negligible, in terms of resource consumption, compared to operations in elliptic curves or RSA base operations (power elevation and inversion). Given that the \idemix model does not have, as far as we know, an implementation on EC, we will focus on the memory usage. In the rest of this section, we consider a credential of $n$ $attributes$ to sign and verify. Results presented below are based on papers \cite{camenisch2007direct}\cite{paquin2011u}\cite{camenisch2002design}\cite{vullers2013efficient}\cite{alpar2015attribute} and simplified schemes presented by Alp{\'a}r Gergely\cite{gergelyalparu-idemix2014}\cite{gergelyalparu-prove2016}.

\subsection{Operations over curve comparison}
\label{label-complexity-operation}
\hspace{3mm}
In this first part of comparative study, we focus on \uprove and \itwopa schemes regarding operations on the elliptic curve. We do not consider the hash function because its fundamental property is that it should be very easy and quick to compute. The comparison results are recorded in "\textbf{table} \ref{label-table-operaion}".

\begin{center} 
\captionof{table}{Operations over curve analysis}
\label{label-table-operaion}
\begin{tabular}{|l|l|c|r|}
\hline
\multicolumn{1}{|p{1.5cm}|}{\centering \textbf{Protocol }} 
& \multicolumn{1}{|p{3cm}|}{\centering \textbf{\uprove } } 
& \multicolumn{1}{|p{2cm}|}{\centering \textbf{\itwopa } } \\
\hline
 \hline
\multicolumn{1}{|p{1.5cm}|}
{\centering 
Issuance
} 
& 

\multicolumn{1}{|p{3cm}|}
{\centering 
$(n+6).M_s+(n+4).A_p$ 
} 
&
\multicolumn{1}{|p{2cm}|}
{\centering 
$(n+6).M_s+2.A_p$ 
} 
\\
\hline
\multicolumn{1}{|p{1.5cm}|}
{\centering 
Verification
} 
&
\multicolumn{1}{|p{3cm}|}
{\centering 
$2.M_s+2.A_p$ 
}
&
\multicolumn{1}{|p{2cm}|}
{\centering 
$2.M_s+1.A_p$ 
} 
\\
\hline
\end{tabular}
\end{center}  
\vspace{1.5mm}

Table \ref{label-table-operaion} shows that, what should be the number of attributes to issue and verify, \uprove  performs more operations in the curve than \itwopap In addition, the issuance is far more expensive in \uprove than in \itwopa especially when the number of attributes grows. We can safely conclude, without going wrong, that \itwopa is more efficient than \uprove in term of computing time when system's parameters are the same.

\subsection{Memory usage comparison}
\label{label-complexity-memory}
\hspace{3mm}
In this second part of comparative study, we  are interested in memory usage. We  compare the number of bits in the issuing phase. This choice is justified by the fact that, in the verification phase, the verifier has nothing to store. We adopt the following notation :

\begin{itemize}
\item {T : Total of variables needed in this phase}
\item {P: Total of variables to be stored permanently in this phase}
\end{itemize}

In addition, we  assume that all variables of a protocol have the same size and this size is the same that the security level. If in the \idemix model we work with 1024 bits, then in \itwopa and \uprove we can work with 160 bits expecting the same security level \cite{sinha2013performance}. Results obtained are recorded in "\textbf{table} \ref{label-table-memory}".

\begin{center} 
\captionof{table}{Memory usage analysis}
\label{label-table-memory}
\begin{tabular}{|l|l|c|r|}
\hline
\multicolumn{1}{|p{1.5cm}|}{\centering \textbf{Protocol }} 
& \multicolumn{1}{|p{1.5cm}|}{\centering \textbf{\idemix } } 
& \multicolumn{1}{|p{1.5cm}|}{\centering \textbf{\uprove } } 
& \multicolumn{1}{|p{1.5cm}|}{\centering \textbf{\itwopa } } \\
\hline
 \hline
\multicolumn{1}{|p{1.5cm}|}
{\centering 
User
} 
& 
\multicolumn{1}{|p{1.5cm}|}
{\centering 
T : n+6\\
P : n+4 
} 
&
\multicolumn{1}{|p{1.5cm}|}
{\centering 
T : n+9\\
P : n+4 
} 
&
\multicolumn{1}{|p{1.5cm}|}
{\centering 
T : n+9\\
P : n+4 
} 
\\
\hline
\multicolumn{1}{|p{1.5cm}|}
{\centering 
Issuer
} 
&
\multicolumn{1}{|p{1.5cm}|}
{\centering 
T : n+9\\
P : n+4 
} 
& 
\multicolumn{1}{|p{1.5cm}|}
{\centering 
T : 2n+8\\
P : 2n+6 
} 
&
\multicolumn{1}{|p{1.5cm}|}
{\centering 
T : 10\\
P : 7
} 
\\
\hline
\end{tabular}
\end{center}

\vspace{1.5mm}
This table show that, at user side, \itwopa and \uprove require around 1024(n+9) bits while \idemix requires around 1024(n+6) bits. However, at issuer side, \itwopa presents interesting results (around 1600 bits  independent on the number of attributes) compared to  \uprove  and \itwopa that require respectively 160(2n+8) and 1024(n+9). We can also see that if implementation is based on  RSA cryptosystem, \uprove presents less interesting results than \idemixp Finally, we can conclude that \itwopa is very efficient compared to \idemix and \uprovev the two leaders in terms of anonymous credentials.

\subsection{Feature comparisons}
\label{label-complexity-feature}
\hspace{3mm}
In this last part of comparative study, we focus on the number of key features that need to be carefully considered when tackling privacy concerns in an IoT context. We consider three scenarios for a feature. It can be totally, partially or not at all supported (not provided). The following legend is adopted:

\begin{itemize}
\item {$\checkmark$ : Fully supported}
\item {$\times$ : Not supported}
\item {$\oslash$ : Partially supported}
\end{itemize}
The "\textbf{table} \ref{label-table-feature}" is a summary of features comparison.

\begin{center} 
\captionof{table}{Features analysis}
\label{label-table-feature}
\begin{tabular}{|l|l|c|r|}
\hline
\multicolumn{1}{|p{3cm}|}{\centering \textbf{Protocol }} 
& \multicolumn{1}{|p{1cm}|}{\centering \textbf{\idemix } } 
& \multicolumn{1}{|p{1cm}|}{\centering \textbf{\uprove } } 
& \multicolumn{1}{|p{1cm}|}{\centering \textbf{\itwopa } } \\
\hline
\hline
\multicolumn{1}{|p{3cm}|}
{\centering 
fully blind signature
} 
& 
\multicolumn{1}{|p{1cm}|}
{\centering 
$\oslash$
} 
&
\multicolumn{1}{|p{1cm}|}
{\centering 
$\checkmark$
} 
&
\multicolumn{1}{|p{1cm}|}
{\centering 
$\checkmark$
} 
\\
\hline
\multicolumn{1}{|p{3cm}|}
{\centering 
Selective disclosure
} 
& 
\multicolumn{1}{|p{1cm}|}
{\centering 
$\checkmark$
} 
&
\multicolumn{1}{|p{1cm}|}
{\centering 
$\checkmark$
} 
&
\multicolumn{1}{|p{1cm}|}
{\centering 
$\checkmark$
} 
\\
\hline
\multicolumn{1}{|p{3cm}|}
{\centering 
Randomization
} 
& 
\multicolumn{1}{|p{1cm}|}
{\centering 
$\checkmark$
} 
&
\multicolumn{1}{|p{1cm}|}
{\centering 
$\times$
} 
&
\multicolumn{1}{|p{1cm}|}
{\centering 
$\checkmark$
} 
\\
\hline
\multicolumn{1}{|p{3cm}|}
{\centering 
Untraceability
} 
& 
\multicolumn{1}{|p{1cm}|}
{\centering 
$\checkmark$
} 
&
\multicolumn{1}{|p{1cm}|}
{\centering 
$\oslash$
} 
&
\multicolumn{1}{|p{1cm}|}
{\centering 
$\checkmark$
} 
\\
\hline
\multicolumn{1}{|p{3cm}|}
{\centering 
Unlinkability
} 
& 
\multicolumn{1}{|p{1cm}|}
{\centering 
$\checkmark$
} 
&
\multicolumn{1}{|p{1cm}|}
{\centering 
$\checkmark$
} 
&
\multicolumn{1}{|p{1cm}|}
{\centering 
$\checkmark$
} 
\\
\hline
\multicolumn{1}{|p{3cm}|}
{\centering 
Unforgeability
} 
& 
\multicolumn{1}{|p{1cm}|}
{\centering 
$\checkmark$
} 
&
\multicolumn{1}{|p{1cm}|}
{\centering 
$\checkmark$
} 
&
\multicolumn{1}{|p{1cm}|}
{\centering 
$\checkmark$
} 

\\
\hline
\multicolumn{1}{|p{3cm}|}
{\centering 
small keys size
} 
& 
\multicolumn{1}{|p{1cm}|}
{\centering 
$\times$
} 
&
\multicolumn{1}{|p{1cm}|}
{\centering 
$\checkmark$
} 
&
\multicolumn{1}{|p{1cm}|}
{\centering 
$\checkmark$
} 
\\
\hline
\multicolumn{1}{|p{3cm}|}
{\centering 
small devices efficiency
} 
& 
\multicolumn{1}{|p{1cm}|}
{\centering 
$\times$
} 
&
\multicolumn{1}{|p{1cm}|}
{\centering 
$\checkmark$
} 
&
\multicolumn{1}{|p{1cm}|}
{\centering 
$\checkmark$
} 
\\
\hline
\multicolumn{1}{|p{3cm}|}
{\centering 
Bandwidth saving 
} 
& 
\multicolumn{1}{|p{1cm}|}
{\centering 
$\times$
} 
&
\multicolumn{1}{|p{1cm}|}
{\centering 
$\checkmark$
} 
&
\multicolumn{1}{|p{1cm}|}
{\centering 
$\checkmark$
} 
\\
\hline
\end{tabular}
\end{center} 
\vspace{1.5mm}

 \idemix  issuance is not totally blind because two of the three parts of the resulting credential are known by the issuer. These two elements are $A$ and $e$ of the signature $(A, e, v)$. In addition, this model is not optimal with low-resource devices and bandwidth optimization. \uprovev meanwhile, does not support the randomization feature. Even if the signature phase is blind, the issues related to traceability are not fully taken into account. Indeed, the triple $ (h', c', r') $ where $ (c', r ') $ constitutes the signature, if presented several times, may be traceable. The problem can be seen in two angles. First $h'$ constituting the user's public key, is required in the verification phase. Since the latter only depends on the signed attributes and the secret key of the user, then the probability of having two credentials with the same public key is almost zero. Secondly, lack of randomization may be a problem. To guarantee non-traceability in \uprovev one needs several credentials\cite{hanzlik2014short}. In our schema, the triple $ (h, R, c) $ can become $ (h, R', c') $ where $ (R', c') $ is a random version of $ (R, c) $ without invalidating the signature.


\section{Conclusion}
\label{label-conclusion}
\hspace{3mm}
Until a recent period, authentication without identification was impossible. Anonymous credentials are a suitable way to do it. In this paper, we propose an efficient Attribute-Based Credentials scheme for IoT. We use Elliptic Curves, Zero Knowledge Proof, Blind Signing, Selective Disclosure and Randomization. Our scheme guarantees anonymity; a fundamental aspect for privacy preserving. As stated in section \ref{label-theorem-resist-attacks}, our scheme is $(\epsilon^{ \prime} ,t^{ \prime},q_{i},q_{h})$-secure in the sense of unforgeability under chosen message attack in the random oracle model. Complexities presented in section \ref{label-complexity} show that our scheme is very suitable with an IoT environment with severely constrained resources. Future work would include performance study, predicate proof over attributes, inspection and revocation protocols.

\bibliographystyle{IEEEtran}
\bibliography{IEEEabrv,IEEEexample}

\begin{thebibliography}{10}
\providecommand{\url}[1]{#1}
\csname url@samestyle\endcsname
\providecommand{\newblock}{\relax}
\providecommand{\bibinfo}[2]{#2}
\providecommand{\BIBentrySTDinterwordspacing}{\spaceskip=0pt\relax}
\providecommand{\BIBentryALTinterwordstretchfactor}{4}
\providecommand{\BIBentryALTinterwordspacing}{\spaceskip=\fontdimen2\font plus
\BIBentryALTinterwordstretchfactor\fontdimen3\font minus
  \fontdimen4\font\relax}
\providecommand{\BIBforeignlanguage}[2]{{%
\expandafter\ifx\csname l@#1\endcsname\relax
\typeout{** WARNING: IEEEtran.bst: No hyphenation pattern has been}%
\typeout{** loaded for the language `#1'. Using the pattern for}%
\typeout{** the default language instead.}%
\else
\language=\csname l@#1\endcsname
\fi
#2}}
\providecommand{\BIBdecl}{\relax}
\BIBdecl

\bibitem{miranda2015internet}
J.~Miranda, N.~M{\"a}kitalo, J.~Garcia-Alonso, J.~Berrocal, T.~Mikkonen,
  C.~Canal, and J.~M. Murillo, ``From the internet of things to the internet of
  people,'' \emph{IEEE Internet Computing}, vol.~19, no.~2, pp. 40--47, 2015.

\bibitem{ciscoprevision}
\BIBentryALTinterwordspacing
D.~H. Joseph~Bradley, Joel~Barbier. (2013) L'internet of everything, un
  potentiel de 14,4 trillions de dollars. [Online]. Available:
  \url{https://www.cisco.com/web/FR/tomorrow-starts-here/pdf/ioe_economy_report_fr.pdf}
\BIBentrySTDinterwordspacing

\bibitem{mbaye2014lightweight}
A.~Mbaye, A.~A. Ciss, and O.~Niang, ``A lightweight identification protocol for
  embedded devices,'' \emph{arXiv preprint arXiv:1408.5945}, 2014.

\bibitem{galparinproceedings}
G.~Alpar and J.-H. Hoepman, ``A secure channel for attribute-based credentials:
  [short paper],'' 11 2013, pp. 13--18.

\bibitem{paquin2011u}
C.~Paquin and G.~Zaverucha, ``U-prove cryptographic specification v1. 1,''
  \emph{Technical Report, Microsoft Corporation}, 2011.

\bibitem{camenisch2002design}
J.~Camenisch and E.~Van~Herreweghen, ``Design and implementation of the idemix
  anonymous credential system,'' in \emph{Proceedings of the 9th ACM conference
  on Computer and communications security}.\hskip 1em plus 0.5em minus
  0.4em\relax ACM, 2002, pp. 21--30.

\bibitem{camenisch2001efficient}
J.~Camenisch and A.~Lysyanskaya, ``An efficient system for non-transferable
  anonymous credentials with optional anonymity revocation,'' in
  \emph{International Conference on the Theory and Applications of
  Cryptographic Techniques}.\hskip 1em plus 0.5em minus 0.4em\relax Springer,
  2001, pp. 93--118.

\bibitem{brands2000rethinking}
S.~A. Brands, \emph{Rethinking public key infrastructures and digital
  certificates: building in privacy}.\hskip 1em plus 0.5em minus 0.4em\relax
  Mit Press, 2000.

\bibitem{irma2016}
\BIBentryALTinterwordspacing
R.~University. (2016) About irma. [Online]. Available:
  \url{https://privacybydesign.foundation/irma-en/}
\BIBentrySTDinterwordspacing

\bibitem{vullers2013efficient}
P.~Vullers and G.~Alp{\'a}r, ``Efficient selective disclosure on smart cards
  using idemix,'' in \emph{IFIP Working Conference on Policies and Research in
  Identity Management}.\hskip 1em plus 0.5em minus 0.4em\relax Springer, 2013,
  pp. 53--67.

\bibitem{camenisch2004signature}
J.~Camenisch and A.~Lysyanskaya, ``Signature schemes and anonymous credentials
  from bilinear maps,'' in \emph{Annual International Cryptology
  Conference}.\hskip 1em plus 0.5em minus 0.4em\relax Springer, 2004, pp.
  56--72.

\bibitem{hanzlik2014short}
L.~Hanzlik and K.~Kluczniak, ``A short paper on how to improve u-prove using
  self-blindable certificates,'' in \emph{International Conference on Financial
  Cryptography and Data Security}.\hskip 1em plus 0.5em minus 0.4em\relax
  Springer, 2014, pp. 273--282.

\bibitem{alpar2012designated}
G.~Alp{\'a}r, L.~Batina, and W.~Lueks, ``Designated attribute-based proofs for
  rfid applications,'' in \emph{International Workshop on Radio Frequency
  Identification: Security and Privacy Issues}.\hskip 1em plus 0.5em minus
  0.4em\relax Springer, 2012, pp. 59--75.

\bibitem{de2017assessment}
J.~De~Fuentes, L.~Gonz{\'a}lez-Manzano, J.~Serna-Olvera, and F.~Veseli,
  ``Assessment of attribute-based credentials for privacy-preserving road
  traffic services in smart cities,'' \emph{Personal and Ubiquitous Computing},
  vol.~21, no.~5, pp. 869--891, 2017.

\bibitem{sinha2013performance}
R.~Sinha, H.~K. Srivastava, and S.~Gupta, ``Performance based comparison study
  of rsa and elliptic curve cryptography,'' \emph{International Journal of
  Scientific \& Engineering Research}, vol.~4, no.~5, pp. 720--725, 2013.

\bibitem{bernal2017holistic}
J.~Bernal~Bernabe, J.~L. Hernandez-Ramos, and A.~F. Skarmeta~Gomez, ``Holistic
  privacy-preserving identity management system for the internet of things,''
  \emph{Mobile Information Systems}, vol. 2017, 2017.

\bibitem{alpar2015attribute}
G.~Alp{\'a}r, \emph{Attribute-based identity management:[bridging the
  cryptographic design of ABCs with the real world]}.\hskip 1em plus 0.5em
  minus 0.4em\relax [Sl: sn], 2015.

\bibitem{liu2015performance}
Z.~Liu, H.~Seo, and Q.~Xu, ``Performance evaluation of twisted edwards-form
  elliptic curve cryptography for wireless sensor nodes,'' \emph{Security and
  Communication Networks}, vol.~8, no.~18, pp. 3301--3310, 2015.

\bibitem{koblitz1987elliptic}
N.~Koblitz, ``Elliptic curve cryptosystems,'' \emph{Mathematics of
  computation}, vol.~48, no. 177, pp. 203--209, 1987.

\bibitem{miller1985use}
V.~S. Miller, ``Use of elliptic curves in cryptography,'' in \emph{Conference
  on the Theory and Application of Cryptographic Techniques}.\hskip 1em plus
  0.5em minus 0.4em\relax Springer, 1985, pp. 417--426.

\bibitem{bernstein2007faster}
D.~J. Bernstein and T.~Lange, ``Faster addition and doubling on elliptic
  curves,'' in \emph{International Conference on the Theory and Application of
  Cryptology and Information Security}.\hskip 1em plus 0.5em minus 0.4em\relax
  Springer, 2007, pp. 29--50.

\bibitem{goldwasserMR85}
\BIBentryALTinterwordspacing
S.~Goldwasser, S.~Micali, and C.~Rackoff, ``The knowledge complexity of
  interactive proof-systems (extended abstract),'' in \emph{Proceedings of the
  17th Annual {ACM} Symposium on Theory of Computing, May 6-8, 1985,
  Providence, Rhode Island, {USA}}, 1985, pp. 291--304. [Online]. Available:
  \url{http://doi.acm.org/10.1145/22145.22178}
\BIBentrySTDinterwordspacing

\bibitem{romahulsing2016}
\BIBentryALTinterwordspacing
A.~Hülsing. (2016) Digital signature schemes and the random oracle model.
  [Online]. Available:
  \url{https://www.win.tue.nl/applied_crypto/2016/20161115_ROM_Signatures.pdf}
\BIBentrySTDinterwordspacing

\bibitem{liu2010efficient}
J.~K. Liu, J.~Baek, J.~Zhou, Y.~Yang, and J.~W. Wong, ``Efficient
  online/offline identity-based signature for wireless sensor network,''
  \emph{International Journal of Information Security}, vol.~9, no.~4, pp.
  287--296, 2010.

\bibitem{jansma2004performance}
N.~Jansma and B.~Arrendondo, ``Performance comparison of elliptic curve and rsa
  digital signatures,'' \emph{nicj. net/files}, 2004.

\bibitem{camenisch2007direct}
J.~Camenisch, ``Direct anonymous attestation explained,'' tech. rep., IBM
  Research, Tech. Rep., 2007.

\bibitem{gergelyalparu-idemix2014}
\BIBentryALTinterwordspacing
G.~ALPAR. (2014) Cryptography fact sheet about idemix’s basic proof
  techniques. [Online]. Available:
  \url{https://privacybydesign.foundation/pdf/idemix_overview.pdf}
\BIBentrySTDinterwordspacing

\bibitem{gergelyalparu-prove2016}
\BIBentryALTinterwordspacing
------. (2016) U-prove cryptography. [Online]. Available:
  \url{http://www.cs.ru.nl/~gergely/objects/u-prove.pdf}
\BIBentrySTDinterwordspacing

\end{thebibliography}

\end{document}